\documentclass[aip,rsi,reprint
]{revtex4-2} 

\usepackage{graphicx}
\usepackage{hyperref}
\hypersetup{colorlinks,allcolors=black}
\begin{document}




\title{Multi-mode Analysis of Surface Losses in a Superconducting Microwave Resonator in High Magnetic Fields}
\author{T. Braine}
  \affiliation{University of Washington, Seattle, WA 98195, USA}
  \email[Correspondence to: ]{tbraine@uw.edu}
\author{G. Rybka}
  \affiliation{University of Washington, Seattle, WA 98195, USA}

\author{A. A. Baker}
  \affiliation{Lawrence Livermore National Laboratory, Livermore, CA 94550, USA}
\author{J. Brodsky}
  \affiliation{Lawrence Livermore National Laboratory, Livermore, CA 94550, USA}
\author{G. Carosi}
  \affiliation{Lawrence Livermore National Laboratory, Livermore, CA 94550, USA}
\author{N. Du}%
  \affiliation{Lawrence Livermore National Laboratory, Livermore, CA 94550, USA}
\author{N. Woollett}
  \affiliation{Lawrence Livermore National Laboratory, Livermore, CA 94550, USA}
  \affiliation{Rigetti Computing, Berkeley, CA 94710, USA}
\author{S. Knirck}
  \affiliation{Fermi National Accelerator Laboratory, Batavia IL 60510, USA}
\author{M. Jones}
  \affiliation{Pacific Northwest National Laboratory, Richland, WA 99354, USA}
  \affiliation{Ozen Engineering Inc, Sunnyvale, CA 94085, USA}
\collaboration{Part of the ADMX Collaboration}
    \homepage[Recent axion search results can be found in \href{https://journals.aps.org/prl/abstract/10.1103/PhysRevLett.127.261803}{PRL 127.261803}\cite{PhysRevLett.127.261803}]{}
    \noaffiliation

\date{\today}

\begin{abstract}
This paper reports on a surface impedance measurement of a niobium titanium superconducting radio frequency (SRF) cavity in a magnetic field (up to $10\,{\rm T}$). A novel method is employed to decompose the surface resistance contributions of the cylindrical cavity end caps and walls using measurements from multiple $TM$ cavity modes. The results confirm that quality factor degradation of a NbTi SRF cavity in a high magnetic field is primarily from surfaces perpendicular to the field (the cavity end caps), while parallel surface resistances (the walls) remain relatively constant. This result is encouraging for applications needing high Q cavities in strong magnetic fields, such as the Axion Dark Matter eXperiment (ADMX), because it opens the possibility of hybrid SRF cavity construction to replace conventional copper cavities.
\end{abstract}


\maketitle 
\section{Motivation}
Superconducting Radio Frequency (SRF) cavity resonators are used in a variety of physics applications to achieve the highest quality factors possible. High quality factor (high Q) cavities are essential for achieving the efficient transfer of large RF power in modern particle accelerators and they are increasingly important for housing and reading out qubits in the quantum information science (QIS) community \cite{SQMS, CAS02}.  In dark matter detection, they are  essential  for  building  up  resonant  power  to  detect  the  tiniest  signals  created  by  a  potential  dark matter candidate:  the axion \cite{Peccei:1977hh,Weinberg:1977ma,Wilczek:1977pj}. Superconductors are the standard material for achieving high Q cavities and $Q>10^{10}$ are commonly achieved in particle accelerator applications \cite{CAS02}. It has recently been demonstrated that the high Q of SRF cavities can be used to achieve the highest sensitivities possible in dark photon searches \cite{RaphaelSRF}.  However, the quality factor of superconducting cavities is severely degraded in the presence of high magnetic fields due to the breakdown of the Meissner effect \cite{Parks,Tinkham}.  Axion haloscopes require microwave resonators to operate in  magnetic  fields in excess of most common superconductors' critical field,  thus  copper  cavities  have  been  the  standard  so  far \cite{Sikivie1985,Rohlf,Blatt}. The measurements in this paper were made as part of a feasibility study for superconductors to be implemented on future runs of the Axion Dark Matter eXperiment (ADMX) operating above 2 $\mathrm{GHz}$. Because of this, only right cylindrical cavity geometries similar to ADMX are discussed; nonetheless, these techniques could be applied to other cavity geometries.
\section{Introduction}
The quality factor of a resonant cavity mode can be defined in terms of energy or in terms of surface resistance: 
\begin{equation}
    Q_0=\omega\frac{U}{P_\mathrm{d}}=\frac{G_\mathrm{s}}{R_\mathrm{s}},
    \label{eqn:Q_factor}
\end{equation}
where $\omega$ is the angular resonant frequency of the cavity mode, $U$ is the electromagnetic energy stored in the cavity mode, $P_\mathrm{d}$ is the power dissipated through the cavity interior surfaces, $G_\mathrm{s}$ is the geometric factor of the cavity mode, and $R_\mathrm{s}$ is the total surface resistance of the cavity. The geometric factor takes into account the spatial distribution of the field for a resonant mode whereas the surface resistance is independent of the mode chosen (not including frequency dependencies). One can arrive at the expression for geometric factor by writing out the total energy stored and power dissipated in the mode \cite{Jackson}:
\begin{equation}
    U=\frac{\mu_0}{2} \int |\vec{H}|^2 \,dV \,,\   P_\mathrm{d}=\frac{R_\mathrm{s}}{2} \int |\vec{H}|^2 \,dS \
    \label{eqn:CavityEnergy}
\end{equation}
The geometric factor is then a ratio of cavity volume and interior surface integrals of the magnetic field of the cavity mode: 
\begin{equation}
    G_\mathrm{s}=\mu_0\omega \frac{ \int |\vec{H}|^2 \,dV }{ \int |\vec{H}|^2 \,dS }
    \label{eqn:GeoFactor}
\end{equation}
\subsection{ADMX Cavity Considerations}
The Axion Dark Matter eXperiment searches for dark-matter axions using an axion haloscope \cite{Sikivie:1983ip}, which comprises a resonant cylindrical cavity inside a magnetic field. The current configuration, a $136\,\ell$ copper-plated cavity and $7.6\,{\rm T}$ magnet, is similar to the one described in Ref.~\cite{Asztalos:2009yp,ASZTALOS201139}. In the presence of an external magnetic field, axions inside the cavity can convert to photons with frequency $f=E/h$, where $E$ is the total energy of the axion, including the axion rest mass energy, plus a small kinetic energy contribution. If the resultant photon frequency matches that of a resonant mode of the cavity, the photon energy developed within the cavity is enhanced by a factor of $Q_0$. However, the power expected from the conversion of an axion into microwave photons in the ADMX experiment is extremely low, $\mathcal{O}(10^{-23}\,{\rm W})$, requiring an ultra low-noise receiver to detect the photons. The expected power deposited into the cavity is of the form found in Ref. \cite{Sikivie1985} but formatted for the standard operating conditions of ADMX:
\begin{widetext}
\begin{equation}
    P_{\mathrm{axion}}=2.2\cdot10^{-23}~\text{W} (\frac{V}{136~\text{L}})(\frac{B}{7.6~\text{T}})^2(\frac{C}{0.4})\\
    (\frac{g_{\gamma}}{0.36})^2(\frac{\rho_a}{0.45~\text{GeV cm}^{-3}})(\frac{f}{740~\text{MHz}})(\frac{Q_L}{30,000})
\label{Eq:Axion_Power}
\end{equation}
\end{widetext}
\noindent Here $V$ is the volume of the cavity, $B$ is the magnitude of the external magnetic field,  $g_{\gamma}$ is the model dependent axion-photon coupling, $\rho_a$ is the local dark-matter density, $f$ is the frequency of the photon, $Q_L$ is the loaded quality factor of the cavity, and $C$ is the form factor of the cavity mode. In terms of cavity construction, one is mostly concerned with $V$, $Q$, and $C$ at a given $f$. Cavity volume is constrained by the target frequency range because the baseline resonant frequency is set by the radius of the cavity, and the height is fixed by the coil height of the solenoid magnet. Form factor is a measure of the overlap of the external magnetic field with cavity mode electric field \cite{Sikivie1985}; for an empty right cylindrical cavity in a solenoid magnet, the form factor is maximized for the $\mathrm{TM_{010}}$ mode with a value of $\approx 0.69$,
\cite{Stern,Peng2000569} but this value decreases when tuning rods are introduced to about $0.4$. Thus these two factors are mostly fixed by the target frequency. The quality factor remains dominated by the RF surface resistance of the material used. For oxygen-free, annealed copper-plated cavities, $Q_0$ is usually around $160,000$ for a $1\,{\rm GHz}$ resonant cavity; higher frequency cavities will have a smaller $Q_0$. Any increase in this value for future axion searches would mean a proportional increase in potential signal power, thus an increase in the average SNR during an axion search, leading ultimately to proportionally faster data-taking runs.

\subsection{Superconducting cavities}
Superconductors have an intimate relationship with magnetic fields. Superconductors are defined by the property of when cooled below a critical temperature, $T_\mathrm{C}$, the material no longer exhibits an internal electrical resistance; it is superconducting. When materials in this state are exposed to external magnetic flux, surface currents are generated which oppose the external field. This phenomena is called the Meissner effect. The Meissner effect does not cause the field to be completely ejected but instead, the field penetrates the superconductor but only to a very small distance, characterized by a parameter $\lambda$, called the London penetration depth, with field strength decaying exponentially to zero within the bulk of the material. A superconductor with little or no magnetic field within its bulk is said to be in a Meissner state. The Meissner effect breaks down when an external magnetic field becomes too large; this returns the superconductor to a normal state \cite{Tinkham}.

Superconductors can be classified based on how this breakdown occurs. In Type I superconductors, there exists one critical field, $H_c$, above which the superconductor abruptly returns to a normal conducting state completely. In Type II superconductors, there are two critical fields, $H_\mathrm{c1}$ and $H_\mathrm{c2}$. If the applied field is below $H_\mathrm{c1}$, the superconductor repels the field completely. If the applied field is above $H_\mathrm{c2}$, the field penetrates the superconductor completely and it returns to a normal state. In between these two critical values, the superconductor exhibits a mixed state, where normal regions develop increasingly as the applied field approaches $H_\mathrm{c2}$. Type II superconductors are predominantly used for SRF cavities operating in magnetic fields, because their $H_\mathrm{c2}$ can be well above $>10\,{\rm T}$, whereas in Type I superconductors $H_c$ is often order $100\,{\rm mT}$ and rarely exceeds one Tesla \cite{Rohlf,Blatt}. 

These normal regions are called flux vortices or fluxons, because they contain circulating super-currents around a normal, flux-penetrating core, and the amount of magnetic flux within each is quantized. The vortex motion within the superconductor is the dominant source of surface resistivity for superconductors in the mixed state \cite{GittlemanRosenblum,Pompeo}. This is what drives quality factor degradation in SRF cavities; the cavity field exerts a Lorentz force on the fluxons dissipating energy into the vortex motion. In the case of ADMX, this degradation was assumed to be severe enough that copper was chosen over superconductors for the highest Q in an $8\,{\rm T}$ field. 

This vortex motion resistivity, however, should be highly dependent on the angle between the applied field and the superconducting surface; a fluxon whose plane of motion is perpendicular to the external field will be in an unstable equilibrium and exhibit the maximum Lorentz force as a result of a cavity field. Whereas a fluxon with the plane of motion parallel to the external field will be in a semi-stable equilibrium and experience a minimal Lorentz force as the result of the cavity field. Therefore it is expected that surfaces perpendicular to the applied field, with resulting higher total flux penetration, will have a much higher increase in surface resistance than surfaces parallel to the applied field, which should have little to no increase in surface resistance. 
Because of this, it is proposed that in SRF cavities, the Q degradation would predominantly be from the cavity end caps (in the case of a right cylindrical cavity) while the cavity walls' resistance remains unchanged by being parallel to the field.

This phenomenon would imply that one could still increase quality factor over standard copper cavities by constructing hybrid surfaced cavities, such as was done by the QUAX collaboration, where only the cavity walls were plated with niobium titanium (NbTi) and the end caps were copper \cite{QUAX}. There have also been notable results in the context of axion searches using YBCO tape on cavities at the CAPP institute \cite{CAPPYBCO}. More recently, $\mathrm{Nb}_3\mathrm{Sn}$ cavities have also shown an improvement over copper in high magnetic fields \cite{PosenNbSN}. In this study, we show the degradation effect directly with a right cylindrical cavity constructed entirely of NbTi, by decomposing the total surface resistance into its constituent sub-surfaces (end caps and walls). We believe this multi-mode sub-surface decomposition technique to be a novel method of determining varying surface resistances in cavities.

\section{Hardware}
\subsection{Physical Properties Measurement System}
A Quantum Design\texttrademark \ Physical Properties Measurement System (PPMS) located at the Lawrence Livermore National Laboratory (LLNL) was used to achieve the temperatures and fields for this measurement. The PPMS is an exchange gas cryostat with a $26.5\,{\rm mm}$ bore, equipped with a superconducting magnet capable of fields up to $16\,{\rm T}$ applied axially along the bore; the minimum temperature is $1.8\,{\rm K}$. The uniform field region height of the magnet was approximately $40\,{\rm mm}$.
\subsection{Cavity}
\begin{figure*}[htb!]
    \centering
    \includegraphics[width=0.7\linewidth]{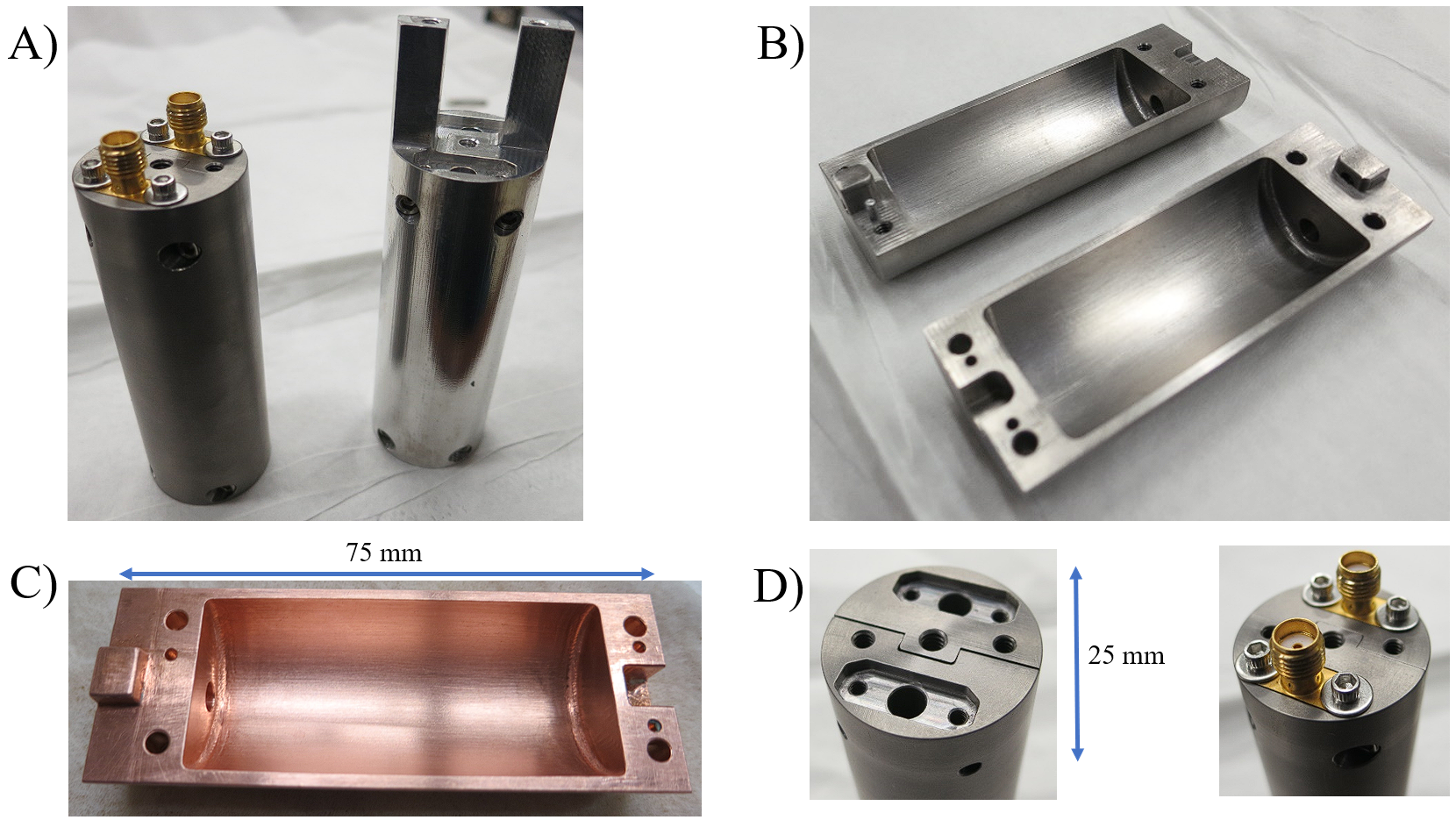}
    \caption{PPMS Clamshell Cavity machined from a bulk NbTi square stock. a) The NbTi cavity assembled next to aluminum prototype b) a profile of both NbTi halves side by side. c) One half of the copper clamshell close-up. d) a close-up of the cavity top without and with fixed pin antennas attached; additional holes are for mounting to probe.}
    \label{fig:NbTiCavity}
\end{figure*}

\begin{figure*}[htb!]
    \centering
    \includegraphics[width=0.7 \linewidth]{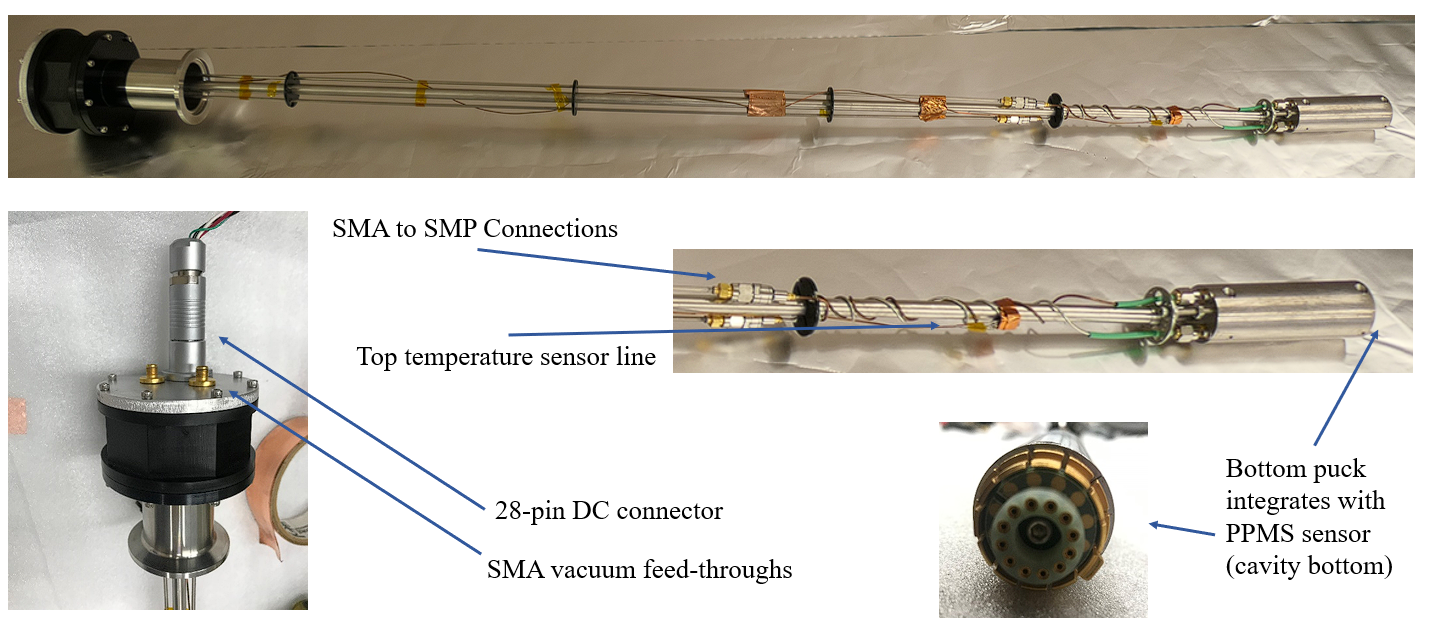}
    \caption{The PPMS readout probe with an aluminum clamshell cavity mounted after a test.  }
    \label{fig:Probe}
\end{figure*}

The cavities used are pictured in Fig.\,\ref{fig:NbTiCavity} and were designed specifically for operating in the PPMS system with an interior diameter of $23\,{\rm mm}$ and interior height of $53\,{\rm mm}$. This height is slightly larger than the 40\,mm uniform field region, but field measurements show that the non-uniformity is a minor perturbation for the $13\,{\rm mm}$ outside the region. This ensured the lowest mode frequencies, $f_{TM_{010}}\approx10\,{\rm GHz}$, for the magnet bore size, $26.5\,{\rm mm}$, and ensured a near uniform magnetic field over the length of the cavity. Additionally, by limiting the height to 55\,mm, the geometric factor of the walls was within an order of magnitude of the total end cap geometric factor (see Table \ref{tab:cavgeofactor}). It is of a 'clamshell' design (See Fig.\,\ref{fig:NbTiCavity}B-C) which lends itself well to being sputtered with superconductor, which is planned to follow on from the work presented in this paper. It has three alignment pin holes and four clamping points to keep the two halves together and aligned. The top contains mounts for two fixed pin antennas, a temperature sensor mount, and two mounting holes for attaching the cavity to a compatible read out probe(Fig.\,\ref{fig:NbTiCavity}D). The readout probe allows for easy insertion into the PPMS bore. The readout probe has two stainless steel SMA cables that run its length to vacuum feed-throughs as well as a 25-pin connector for DC lines, currently only being used for the cavity top temperature sensor (see Fig.\,\ref{fig:Probe}). The bottom contains a center line hole for attaching a Quantum Design\texttrademark \ 'puck' that thermalizes the cavity bottom to the PPMS internal temperature sensor. Versions made of bulk NbTi, aluminum, and copper have been machined so far pictured in Fig.\,\ref{fig:NbTiCavity}A-C. The aluminum prototype was used for fitting and troubleshooting the design, before the final design was machined out of copper and NbTi.
\begin{table}
    \centering
 \begin{tabular}{||c|| c| c |c||} 
 \hline
  & $TM_{010}$ & $TM_{011}$ & $TM_{012}$ \\ [0.5ex] 
 \hline\hline
 Walls & $457.0\pm2.2$ & $494.6\pm3.4$ & $549.9\pm5.5$ \\ 
 \hline
 Top End Cap & $3846\pm41$ & $2045\pm15$ & $2214\pm6$ \\
 \hline
 Bottom End Cap & $4127\pm55$ & $2029\pm26$ & $2192\pm21$ \\
 \hline
 Total Endcaps & $1991\pm38$ & $1019\pm17$ & $1101\pm12$ \\
 \hline
 Total & $371.7\pm9.3$ & $332.9\pm7.0$ & $366.8\pm6.2$ \\ [1ex] 
 \hline
    \end{tabular}
    \caption{Geometric factors for clamshell cavity design for first three $TM$ modes by interior surfaces. These were calculated from simulation in the HFSS software package and are expressed in Ohms. }
    \label{tab:cavgeofactor}
\end{table}
\begin{figure*}[htb!]
    \centering
    \includegraphics[width=0.6\linewidth]{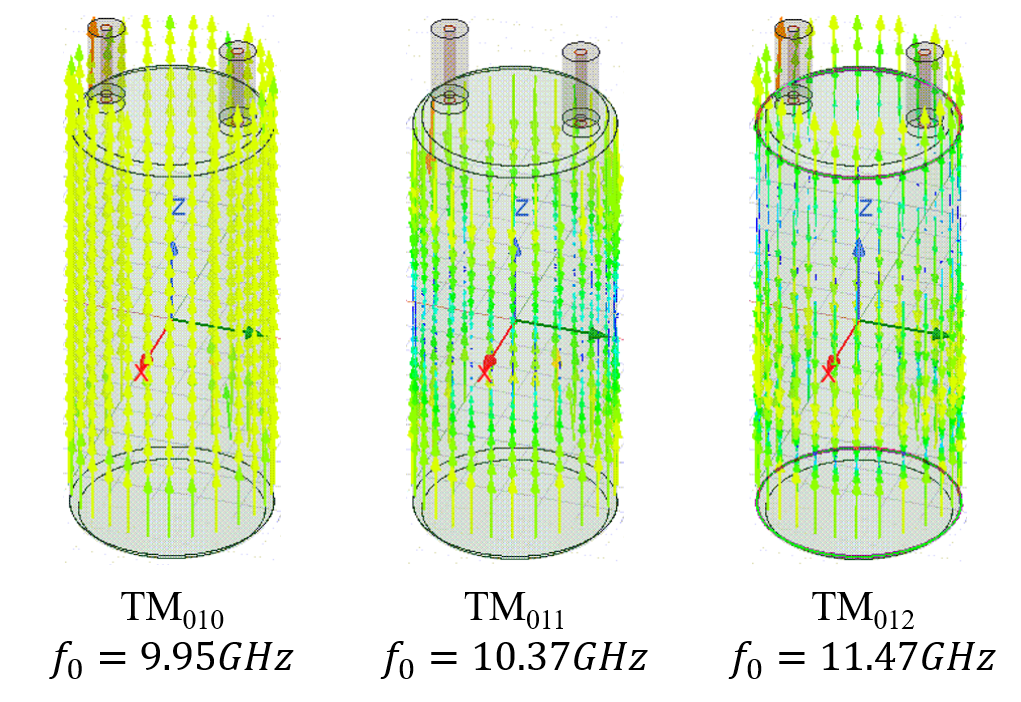}
    \caption{A field overlay of the modes' surface currents in Ansys\texttrademark HFSS software}
    \label{fig:CavityCurrents}
\end{figure*}

\section{Methods}
\subsection{Decomposing into Sub-surface Resistances}
The previously mentioned definition of geometric factor can be used to define geometric factors for each subsurface, $S_\mathrm{n}$ of the total interior cavity surface, $S$, where $S=S_1+S_2+...+S_N$:
\begin{equation}
    G_\mathrm{n}=\mu_0\omega \frac{ \int |\vec{H}|^2 \,dV }{ \int |\vec{H}|^2 \,dS_n }
    \label{eqn:SubGeoFactor}
\end{equation}
This implies that subsurface geometric factors add reciprocally:
\begin{equation}
    \frac{1}{G_\mathrm{tot}}=\frac{1}{G_1}+\frac{1}{G_2}+...+\frac{1}{G_\mathrm{N}}
    \label{eqn:GeoFactorAdd}
\end{equation}
We can also define an associated subsurface quality factor, $Q_\mathrm{n}$, which also follows a reciprocal relation:
\begin{equation}
    \frac{1}{Q_\mathrm{tot}}=\frac{1}{Q_1}+\frac{1}{Q_2}+...+\frac{1}{Q_\mathrm{N}}
    \label{eqn:QAdd}
\end{equation}
Note that because of the mode structure interacting with sub-surfaces differently, $R_\mathrm{s_{tot}} \neq R_{s_1}+R_{s_2}+...+R_\mathrm{s_N}$, rather the total mode resistance will be the weighted sum of the constituent surface resistances. By substituting equations \ref{eqn:GeoFactorAdd} and \ref{eqn:QAdd}, into equation \ref{eqn:Q_factor}, and solving for $R_\mathrm{tot}$ we arrive at an equation of the form, $R_\mathrm{s_{tot}}=C_1R_{s_1}+C_2R_{s_2}+...+C_\mathrm{N}R_\mathrm{s_N}$, where the 'weights' are defined as:
\begin{equation}
    C_\mathrm{n}=\frac{\frac{1}{G_\mathrm{n}}}{\sum_{i=1}^{N}\frac{1}{G_\mathrm{i}}}
    \label{eqn:Rweights}
\end{equation}
In the simplest case, $S=S_1+S_2$, the weights are: 
\begin{equation}
    C_\mathrm{1}=\frac{G_\mathrm{2}}{G_\mathrm{1}+G_\mathrm{2}} ,\ C_\mathrm{2}=\frac{G_\mathrm{1}}{G_\mathrm{1}+G_\mathrm{2}}
    \label{eqn:Rweights2x2}
\end{equation}

\subsection{Converting Multiple Mode Resistances to Surface Resistances}
Using this mode decomposition we can then implement a novel method for measuring RF surface impedance; by measuring multiple cavity mode quality factors, and simulating the modes' geometric factors, the resistance contributions of sub-surfaces can be calculated. The geometric factor for a given cavity mode, $m$, and given subsurface, $s$, is denoted $G_{ms}$. Following the previous analysis we write the weighted sum that gives the total resistance for mode $m$:
\begin{equation}
    R_\mathrm{m}=C_{\mathrm{m}1}R_{1}+C_{\mathrm{m}2}R_{2}+...+C_\mathrm{mN}R_\mathrm{N}
    \label{eqn:ModeResistanceDecomp}
\end{equation}
Note that in this case the surface resistances are independent of mode. This linear combination implies a matrix equation: 
\begin{equation}
    \vec{R_\mathrm{m}}=C_\mathrm{ms}\vec{R_\mathrm{s}}
    \label{eqn:Cmatrix}
\end{equation}
where $\vec{R_\mathrm{m}}$ is a list of total mode resistances corresponding to $\frac{G_\mathrm{m}}{Q_\mathrm{m}}$, the total geometric and total quality factor for the mode. The surface resistance vector, $\vec{R_\mathrm{s}}$, is a list of surface resistances for each subsurface independent of mode. If these two vectors are equal in length, or rather $\vec{R_\mathrm{m}}$ is at least as long as the desired amount of sub surfaces one wants to decompose to, a rectangular matrix can be formed, and the matrix equation \ref{eqn:Cmatrix} can be inverted:
\begin{equation}
    \vec{R_\mathrm{s}}=C_\mathrm{ms}^{-1}\vec{R_\mathrm{m}}
    \label{eqn:Cinvmatrix}
\end{equation}
In short, a set of N mode quality factor measurements yields of decomposition up to N subsurface resistances. By taking additional mode measurements one can preform multiple decompositions using different combinations of modes to yield better statistics. It is important to note, however, that this only works for mode combinations that yield a non-zero determinant. In the 2x2 case, the determinant is zero when $G_{1i}=G_{2i}$, or in other words, when the modes do not have distinct geometric factors. Therefore, it is best to choose modes that have distinct field overlaps with the surfaces of interest, while also being close enough in frequency that any frequency dependencies are below the error threshold. Table \ref{tab:determinants} shows the determinant for the 3 possible mode combinations and their associated error; error estimates will be discussed later in this paper.
\begin{table}
    \centering
 \begin{tabular}{||c|| c| c||} 
 \hline
  Mode Combination & $\mathrm{Det(C_{ms})})$ & $\mathrm{\Delta(Det(C_{ms}))}$ \\ [0.5ex] 
 \hline\hline
 $TM_{010}/TM_{011}$ & 0.140 & 0.016  \\ 
 \hline
 $TM_{010}/TM_{012}$ & 0.146 & 0.016  \\
 \hline
 $TM_{011}/TM_{012}$ & 0.006 & 0.013  \\
 \hline
    \end{tabular}
    \caption{Determinant of Weight Matrix and associated error for the three possible mode combinations. }
    \label{tab:determinants}
\end{table}

\subsection{Calculating Geometric factor and weights}
The Ansys HFSS software package was used to simulate the clamshell cavity design pictured in Fig.\,\ref{fig:NbTiCavity}. The simulated cavity modes' field structures and surface currents (Fig.\,\ref{fig:CavityCurrents}) were used to calculate the volume and surface integrals shown in equation \ref{eqn:GeoFactor} to arrive at a set of nine geometric factors corresponding to the three modes measured ($TM_{010}$, $TM_{011}$, and $TM_{012}$) and the three sub-surfaces we wished to calculate (the interior walls, bottom end cap, and top end cap). These geometric factors are listed in Table \ref{tab:cavgeofactor}. It is important to note the distinction between top and bottom end cap, because the top caps contain the antenna ports; as one can see in Table \ref{tab:cavgeofactor} this doesn't make a big difference, but it is larger than the uncertainty in the geometric factor. Software was developed that could take these results and construct the weighted matrix for any combination of modes and sub-surfaces. One of these software tools allowed us to combine geometric factors of the top and bottom end cap surfaces into a total end cap surface geometric factor, as well as the total mode geometric factor.
\subsection{Measuring Quality factor}
The cavity mode quality factor can be measured \textit{in situ} by using a vector network analyzer (VNA) to measure the scattering parameters through the antennas mounted to the cavity top (see Fig.\,\ref{fig:NbTiCavity}D). Transmission measurements ($S_{21}$) are taken by measuring the proportion of power transmitted between two antennae mounted in the cavity. The quality factor is measured as the full-width at half maximum (FWHM) of the Lorentzian observed in the transmission measurement:
\begin{equation}
    Q_\mathrm{L}=\frac{f_0}{\Delta f}
    \label{eqn:QL}
\end{equation}
where $f_0$ is the mode resonant frequency and $\Delta f$ is the FWHM of the peak. This equation is equally accurate for reflection dips, however it tends to be less consistent for a cavity with weakly coupled antennae that produces very small reflection dips. The coupling coefficient of the antennae is measured through the reflection parameters and is defined by the following equation:
\begin{equation}
    \beta = \frac{1+\mathrm{sign}(\angle \Gamma_\mathrm{cav}(f_0)-\pi)|\Gamma_\mathrm{cav}(f_0)|}{1-\mathrm{sign}(\angle \Gamma_\mathrm{cav}(f_0)-\pi)|\Gamma_\mathrm{cav}(f_0)|}
    \label{eqn:beta}
\end{equation}
where $\angle \Gamma_\mathrm{cav}(f_0)$ is the cavity reflection coefficient phase on resonance and $|\Gamma_\mathrm{cav}(f_0)|$ is the cavity reflection coefficient magnitude on resonance. To obtain the cavity reflection coefficient, a linear fit is performed on the edges of the VNA scan (off-resonance data) to obtain the phase delay of the coaxial cables from VNA to antennas, and this phase delay is subtracted from the VNA reflection data to estimate the cavity reflection coefficient on resonance. The off-resonance edge regions are defined to be the continuous set of points from either edge that is less than $\frac{1}{3}$ the peak height of the Lorentzian peak; this fraction is arbitrary but needs to be low enough to exclude resonance data. The unloaded or intrinsic cavity quality factor is then defined to be: 
\begin{equation}
    Q_\mathrm{0}=Q_\mathrm{L}(1+\beta)
    \label{eqn:Q0}
\end{equation}
Since the cavity measurement was performed with fixed pin length antennae, the antennae were cut back to be as weakly coupled as possible while still having an observable reflection dip. This technique minimized the $\beta$ coefficient, while still being measurable, so that the loaded quality factor was very close to the unloaded value. We estimate $\beta <10^{-2}$ for the cavity assembly. A more in depth discussion of quality factor measurement can be found in Ref. \cite{Brubaker}.

\section{Results}
\begin{figure*}[htb!]
    \centering
    \includegraphics[width=0.5\linewidth]{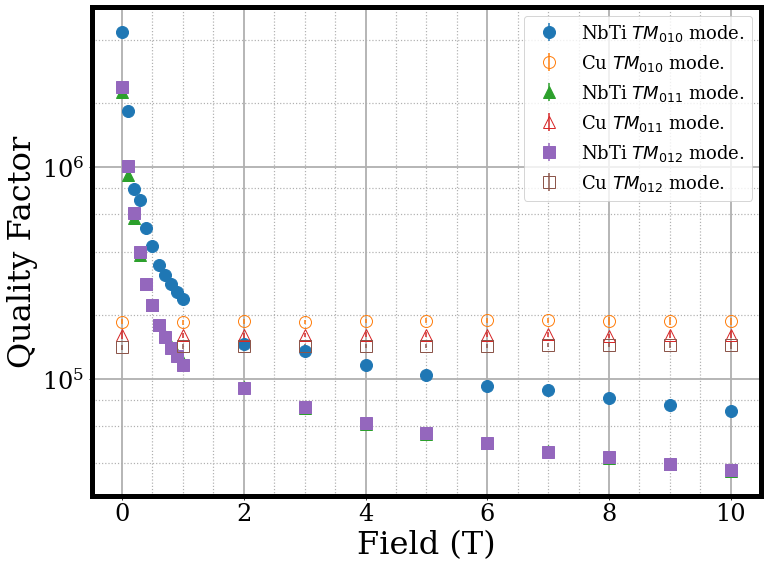}
    \caption{The cavity quality factor for the first three $TM$ modes as function of applied magnetic field. Data was taken at a temperature of $2 \ K$. }
    \label{fig:ModeQvsB}
\end{figure*}
\begin{figure*}[htb!]
    \centering
    \includegraphics[width=0.5\linewidth]{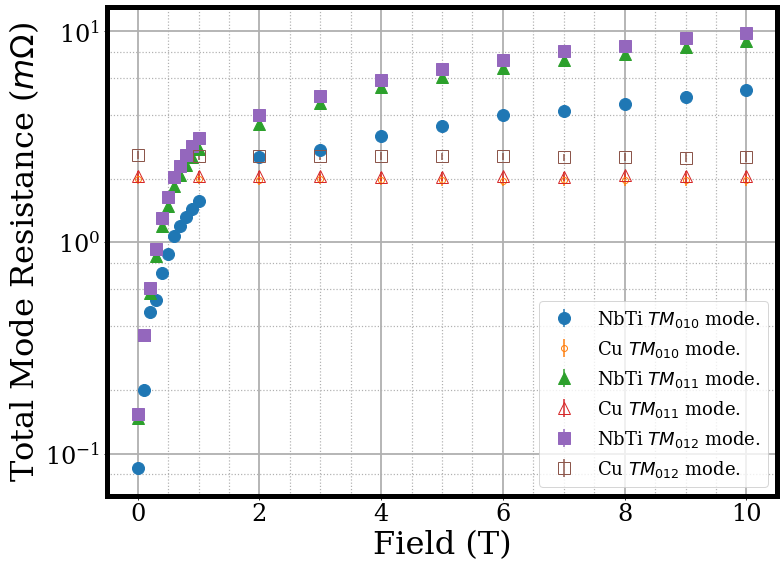}
    \caption{The total mode surface resistance for the first three $TM$ modes as function of applied magnetic field. }
    \label{fig:ModeRvsB}
\end{figure*}
\begin{figure*}[htb!]
    \includegraphics[width=0.5\linewidth]{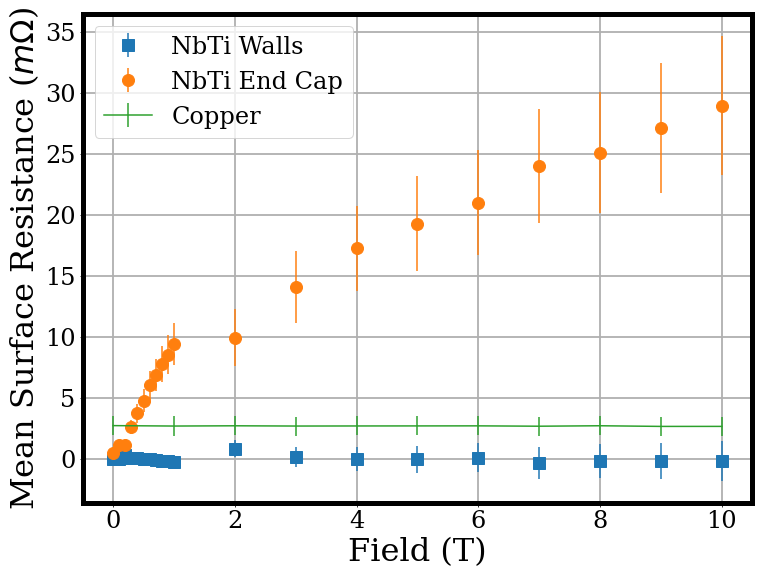}
    \caption{The mean surface resistance contributions of the walls and end caps of the NbTi cavity based on two different 2-mode decompositions: $TM_{010}$/$TM_{011}$ and $TM_{010}$/$TM_{012}$. This is compared to the mean copper cavity resistance, averaged over surface type for the same mode combinations. }
    \label{fig:TM010RSvsB}
\end{figure*}
The clamshell cavity was cooled to $2\,{\rm K}$ in zero field by cooling to $10\,{\rm K}$ at $10\,{\rm K/min}$, then to $2\,{\rm K}$ at $1\,{\rm K/min}$ and then held for $\approx\,10$ minutes to allow the cavity to fully thermalize. The magnet was then ramped in steps ($0.1\,{\rm T}$ steps from $0$ to $1\,{\rm T}$ and $1\,{\rm T}$ steps from 1 to $10\,{\rm T}$) , and held in between until the mode frequencies and Q factors settled to constant values ($\approx\,1$ minute), at which point the resonant peaks were recorded from the VNA for the three $TM$ modes of interest ($TM_{010}$, $TM_{011}$, and $TM_{012}$). The zero field Q factor for the NbTi cavity was measured to be $\approx\,70,000$ at $2\,{\rm K}$ directly after the cavity was machined. The cavity was then sent to Jefferson Laboratory for a cleaning and annealing process. Post-annealing, the cavity had a zero field quality factor at $2\,{\rm K}$ of nearly 1 million.
For comparison, the copper cavity had a Q of about 180,000 at $2\,{\rm K}$. The critical temperature of the NbTi was observed to be approximately $\approx\,8\,{\rm K}$ based on the beginning of the superconducting transition for both the pre-annealed and post-annealed cavity. The mode quality factor as a function of applied magnetic field is shown in Fig.\,\ref{fig:ModeQvsB}. These quality factors were then converted to total mode resistances via equation \ref{eqn:Q_factor} shown in Fig. \ref{fig:ModeRvsB}.
\par In order to test the consistency of this method, a two mode decomposition to the walls and total end caps contributions was performed with combinations of the three modes measured. The $TM_{010}$/$TM_{011}$ and $TM_{010}$/$TM_{012}$ decomposition surface resistances agreed very well, and confirmed the hypothesis that the end caps were the primary contribution to the surface resistance in field (Fig.\,\ref{fig:TM010RSvsB}). The $TM_{011}$/$TM_{012}$ combination unfortunately did not produce a physical result; this is because the geometric factors of the two modes were nearly identical, making the determinant of the weights matrix very nearly zero as discussed earlier (almost two orders of magnitude smaller than the other two combinations), inflating errors when taking the inverse (see Table \ref{tab:determinants}). Thus, this mode combination was excluded from the data set.  

\subsection{Error Budget}
\begin{figure*}[htb!]
    \includegraphics[width=0.5\textwidth]{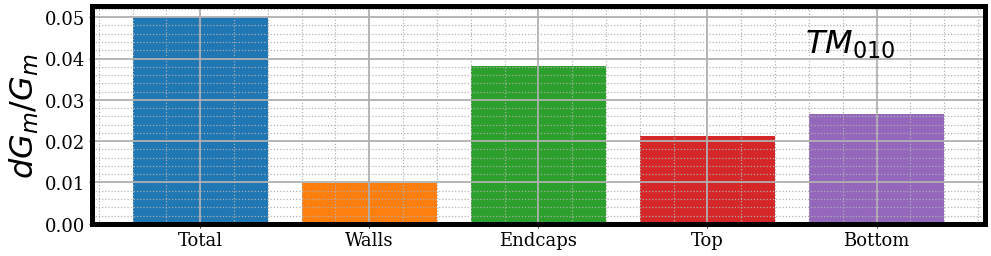}
    \includegraphics[width=0.5\textwidth]{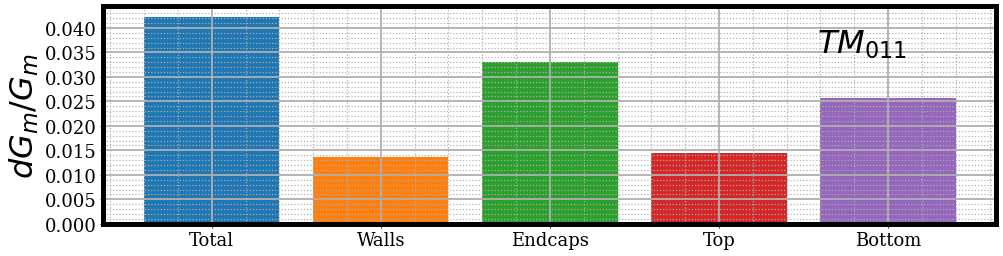}
    \includegraphics[width=0.5\textwidth]{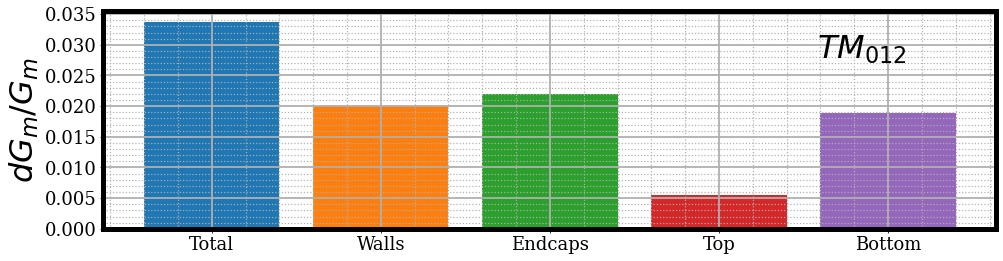}
    \caption{The estimated fractional error in geometric factor by mode and subsurface. Note that the end caps factor is the sum of the top and bottom end cap contributions, and the total geometric factor is the sum of the end caps and walls.}
    \label{fig:dGBargraph}
\end{figure*}
The main two sources of error in this resistance calculation are the uncertainties in the geometric factor from simulation and quality factor from the VNA fitting routine. In terms of the latter, the uncertainty in the Lorentzian fit parameters is returned by the fitting routine to give an estimate for $\Delta Q$; We used the \verb|Scipy.optimize.curve_fit| function \cite{SciPy}, which returns an estimated covariance matrix for the fit parameters.  Typically, $\Delta Q$ was on the order of $10^2$, and any data points with $\frac{\delta Q}{Q} > 0.10$ were noted as outliers, subsequently reviewed by plotting the data. The seven outliers were for the superconducting cavity at low field. When inspecting the fit it became clear that the fit function failed for a large window compared to the peak width. These fits were corrected by reducing the window size during post-processing.
\newline\indent The uncertainty in the geometric factor is calculated by multiple simulations: by varying the model dimensions in the simulation software based on the machining tolerances that cavity was manufactured to ($+/-0.005"$), and re-running the geometric factor calculation for each mode and surface. This was done using a parametric analysis in HFSS, varying the cavity height and radius, 9 sets of geometric factors were generated around the baseline value of cavity height and radius. These sets are the nine geometric factors corresponding to one simulation geometry. The mean and median of each geometric factor were calculated across the sets to obtain a mean and median set of geometric factors. Each simulation set was then compared to the mean and median sets; the set that was closest to these was chosen as the set of geometric factors to be used and are the values listed in Table \ref{tab:cavgeofactor} (This set happened to be closest to both the mean and median sets). Next, the set that deviated the most from the mean set was used to calculate the uncertainty. $\Delta G_{ms}$ was calculated as the difference between these minimum and maximum deviation sets. This comparison process is necessary because the mean values of all individual geometric factors do not form a set of geometric factors corresponding to a real cavity geometry, therefore combining them together is not going to produce an accurate representation of the cavity. Instead, one must use a set of geometric factors where all the values correspond to the same simulation geometry. The fractional uncertainties in the geometric factors are plotted in Fig.\,\ref{fig:dGBargraph}. From there, using rules of error propagation, we can derive the expressions for the uncertainty in the total mode resistances $R_m$:
\begin{equation}
    |\frac{\Delta R_{\mathrm{m}}}{R_{\mathrm{m}}}|=\sqrt{(\frac{\Delta G_{\mathrm{m}}}{G_{\mathrm{m}}})^2+(\frac{\Delta Q_{\mathrm{m}}}{Q_{\mathrm{m}}})^2}
    \label{eqn:dR_m}
\end{equation}
where $G_m=\sum_{s}G_{ms}$ and $|\Delta G_m|=\sqrt(\sum_{s}\Delta G_{ms}^2$).
We show the uncertainty in the weights, $\Delta C_{ms}$, for the 2x2 case where $C_{ix}=\frac{G_{iy}}{G_{ix}+G_{iy}}$ where if $x=1$, then $y=2$ and vice-versa:
\begin{equation}
    |\frac{\Delta C_{\mathrm{ix}}}{C_{\mathrm{ix}}}|=\sqrt{(\frac{\Delta G_{\mathrm{iy}}}{G_{\mathrm{iy}}})^2+(\frac{\Delta G_{\mathrm{ix}}^2+\Delta G_{\mathrm{iy}}^2}{(\Delta G_{\mathrm{ix}}+\Delta G_{\mathrm{iy}})^2}}
    \label{eqn:dC_ms}
\end{equation}
Taking the inverse of the 2x2 matrix gives us an expression for $\Delta C_{ms}^{-1}$ in terms of error in $C_{ms}$ components as well as error in the determinant of the matrix $\Delta (det(C))$:
\begin{widetext}
\begin{equation}
        |\frac{\Delta (det(C))}{det(C)}|=\sqrt{C_{\mathrm{11}}^2\Delta C_{\mathrm{22}}^2+C_{\mathrm{22}}^2\Delta C_{\mathrm{11}}^2+C_{\mathrm{12}}^2\Delta C_{\mathrm{21}}^2+C_{\mathrm{21}}^2\Delta C_{\mathrm{12}}^2}
\label{eqn:ddetC_ms}
\end{equation}
\end{widetext}
\begin{equation}
        |\frac{\Delta C_{\mathrm{xy}}^{-1}}{C_{\mathrm{xy}}^{-1}}|=\sqrt{(\frac{\Delta C_{\mathrm{xy}}}{C_{\mathrm{xy}}})^2+(\frac{\Delta (det(C))}{det(C)})^2}
        \label{eqn:dCinv_ms}
\end{equation}
It is important to note that if the determinant is near zero because the modes aren't appreciably different, that the fractional error will be inflated after this step. This was the case with the $TM_{011}$ $TM_{012}$ mode combination, where the determinant was an order of magnitude smaller than its error, making it effectively zero, and giving fractional errors $>>1$, therefore we did not include this decomposition in our data set.(See Table \ref{tab:determinants}) Subsequently the fractional error in the $R_{s}$ set of values was then found to be:
\begin{equation}
        |\frac{\Delta R_{\mathrm{s}}}{R_{\mathrm{s}}}|=\sqrt{\sum_{x=1}^{2}R_{x}^{2}(\Delta C_{\mathrm{sx}}^{-1})^2+(C_{\mathrm{sx}}^{-1})^2\Delta R_x^2}
        \label{eqn:dR_s}
\end{equation}
where $x$ is used to specifically refer to the second index of the $C^{-1}$ matrix and the mode number of the total mode resistance, and $s$ refers to the two defined surfaces being decomposed. The note to make here is that since many resistance values will be near zero for superconducting surfaces, the fractional error might be large even though the raw error is relatively constant (for our results $\pm1\,{\rm m\Omega}$). A summary of the maximum error in each factor is shown in Table \ref{tab:errorbudget}.

\begin{table}
\centering
\renewcommand{\arraystretch}{1.3}
\begin{tabular}{@{}lcr@{}}
\toprule
Parameter & Variable & Maximum Uncertainty \\
\hline
Quality Factor & $\Delta Q$ & $4.2\%$ \\
\hline
Geometric Factor & $\Delta G_{ms}$ & $5.0\%$ \\
 \hline
Total Mode Resistance& $\Delta R_{m}$ & $6.6\%$ \\  
 \hline
Weight matrix & $\Delta C_{ms}$ & $4.9\%$ \\  
 \hline
Inverse weight matrix & $\Delta C_{ms}^{-1}$ & $12.6\%$ \\
 \hline
Surface Resistance & $\Delta R_{s}$ & $5.7  m\Omega$ \\ 
 \hline
\end{tabular}
\caption{A Summary of the maximum estimated uncertainties for the measurement setup. They are expressed as percentages of their nominal values with the exception of surface resistance, since many values measured were effectively zero while the cavity was superconducting. Note that $\Delta C_{ms}^{-1}$ and $\Delta R_{s}$ does not include the $TM_{011}$/$TM_{012}$ decomposition because of near-zero determinant.}
\label{tab:errorbudget}
\end{table}

\section{Conclusions}
This result is encouraging for the prospect of developing hybrid SRF cavities for axion detection. Although the end caps surface resistance does increase with magnetic field such that it is greater than copper, the cavity walls parallel to the external magnetic field had a relatively constant surface resistance that was lower than copper over the field range. Based on the highest wall resistance measured with maximum positive error, the minimum improvement factor over an equivalent copper surface was approximately a factor of 2.3. This implies that simply replacing the endcaps with copper would give us a net higher Q  for fields
up to at least $10\,{\rm T}$. By taking our method in reverse, using the maximum NbTi wall resistance with the copper end cap resistance, to solve for a new hybrid total mode resistance, we can project that a hybrid cavity of this construction would then have a minimum Q of 315,000 for the $TM_{010}$ mode throughout the field range; this is actually a factor of 1.67 over an all copper cavity with the difference originating in the weighting of surface contributions. Future work will include coating superconducting films on the walls of a copper version of this cavity as a hybrid prototype and testing the cavity in the PPMS as a proof of concept. 


%
%

%

\begin{acknowledgments}
Special thanks to Robert Rimmer's group at Jefferson Lab for cleaning and annealing the NbTi Cavity. We'd also like to thank former LLNL technician Eric Robertson for machining the NbTi cavity itself. This work was supported by the U.S. Department of Energy through Grants Nos. DE-SC0009723, DE-SC0010296, DE-SC0010280, DE-SC0010280, DE-FG02-97ER41029, DE-FG02-96ER40956, DE-AC52-07NA27344, and DE-C03-76SF00098. This manuscript has been authored by Fermi Research Alliance, LLC under Contract No. DE-AC02-07CH11359 with the U.S. Department of Energy, Office of Science, Office of High Energy Physics. Additional support was provided by the Heising-Simons Foundation and by the LDRD offices of the Lawrence Livermore and Pacific Northwest National Laboratories. LLNL Release Number: LLNL-JRNL-838971.
\end{acknowledgments}

\bibliography{ModeSurfaceDecompBIB.bib}

\begin{thebibliography}{25}%
\makeatletter
\providecommand \@ifxundefined [1]{%
 \@ifx{#1\undefined}
}%
\providecommand \@ifnum [1]{%
 \ifnum #1\expandafter \@firstoftwo
 \else \expandafter \@secondoftwo
 \fi
}%
\providecommand \@ifx [1]{%
 \ifx #1\expandafter \@firstoftwo
 \else \expandafter \@secondoftwo
 \fi
}%
\providecommand \natexlab [1]{#1}%
\providecommand \enquote  [1]{``#1''}%
\providecommand \bibnamefont  [1]{#1}%
\providecommand \bibfnamefont [1]{#1}%
\providecommand \citenamefont [1]{#1}%
\providecommand \href@noop [0]{\@secondoftwo}%
\providecommand \href [0]{\begingroup \@sanitize@url \@href}%
\providecommand \@href[1]{\@@startlink{#1}\@@href}%
\providecommand \@@href[1]{\endgroup#1\@@endlink}%
\providecommand \@sanitize@url [0]{\catcode `\\12\catcode `\$12\catcode
  `\&12\catcode `\#12\catcode `\^12\catcode `\_12\catcode `\%12\relax}%
\providecommand \@@startlink[1]{}%
\providecommand \@@endlink[0]{}%
\providecommand \url  [0]{\begingroup\@sanitize@url \@url }%
\providecommand \@url [1]{\endgroup\@href {#1}{\urlprefix }}%
\providecommand \urlprefix  [0]{URL }%
\providecommand \Eprint [0]{\href }%
\providecommand \doibase [0]{https://doi.org/}%
\providecommand \selectlanguage [0]{\@gobble}%
\providecommand \bibinfo  [0]{\@secondoftwo}%
\providecommand \bibfield  [0]{\@secondoftwo}%
\providecommand \translation [1]{[#1]}%
\providecommand \BibitemOpen [0]{}%
\providecommand \bibitemStop [0]{}%
\providecommand \bibitemNoStop [0]{.\EOS\space}%
\providecommand \EOS [0]{\spacefactor3000\relax}%
\providecommand \BibitemShut  [1]{\csname bibitem#1\endcsname}%
\let\auto@bib@innerbib\@empty
\bibitem [{\citenamefont {Bartram}\ \emph {et~al.}(2021)\citenamefont
  {Bartram}, \citenamefont {Braine}, \citenamefont {Burns}, \citenamefont
  {Cervantes}, \citenamefont {Crisosto}, \citenamefont {Du}, \citenamefont
  {Korandla}, \citenamefont {Leum}, \citenamefont {Mohapatra}, \citenamefont
  {Nitta}, \citenamefont {Rosenberg}, \citenamefont {Rybka}, \citenamefont
  {Yang}, \citenamefont {Clarke}, \citenamefont {Siddiqi}, \citenamefont
  {Agrawal}, \citenamefont {Dixit}, \citenamefont {Awida}, \citenamefont
  {Chou}, \citenamefont {Hollister}, \citenamefont {Knirck}, \citenamefont
  {Sonnenschein}, \citenamefont {Wester}, \citenamefont {Gleason},
  \citenamefont {Hipp}, \citenamefont {Jois}, \citenamefont {Sikivie},
  \citenamefont {Sullivan}, \citenamefont {Tanner}, \citenamefont {Lentz},
  \citenamefont {Khatiwada}, \citenamefont {Carosi}, \citenamefont {Robertson},
  \citenamefont {Woollett}, \citenamefont {Duffy}, \citenamefont {Boutan},
  \citenamefont {Jones}, \citenamefont {LaRoque}, \citenamefont {Oblath},
  \citenamefont {Taubman}, \citenamefont {Daw}, \citenamefont {Perry},
  \citenamefont {Buckley}, \citenamefont {Gaikwad}, \citenamefont {Hoffman},
  \citenamefont {Murch}, \citenamefont {Goryachev}, \citenamefont {McAllister},
  \citenamefont {Quiskamp}, \citenamefont {Thomson},\ and\ \citenamefont
  {Tobar}}]{PhysRevLett.127.261803}%
  \BibitemOpen
  \bibfield  {author} {\bibinfo {author} {\bibfnamefont {C.}~\bibnamefont
  {Bartram}}, \bibinfo {author} {\bibfnamefont {T.}~\bibnamefont {Braine}},
  \bibinfo {author} {\bibfnamefont {E.}~\bibnamefont {Burns}}, \bibinfo
  {author} {\bibfnamefont {R.}~\bibnamefont {Cervantes}}, \bibinfo {author}
  {\bibfnamefont {N.}~\bibnamefont {Crisosto}}, \bibinfo {author}
  {\bibfnamefont {N.}~\bibnamefont {Du}}, \bibinfo {author} {\bibfnamefont
  {H.}~\bibnamefont {Korandla}}, \bibinfo {author} {\bibfnamefont
  {G.}~\bibnamefont {Leum}}, \bibinfo {author} {\bibfnamefont {P.}~\bibnamefont
  {Mohapatra}}, \bibinfo {author} {\bibfnamefont {T.}~\bibnamefont {Nitta}},
  \bibinfo {author} {\bibfnamefont {L.~J.}\ \bibnamefont {Rosenberg}}, \bibinfo
  {author} {\bibfnamefont {G.}~\bibnamefont {Rybka}}, \bibinfo {author}
  {\bibfnamefont {J.}~\bibnamefont {Yang}}, \bibinfo {author} {\bibfnamefont
  {J.}~\bibnamefont {Clarke}}, \bibinfo {author} {\bibfnamefont
  {I.}~\bibnamefont {Siddiqi}}, \bibinfo {author} {\bibfnamefont
  {A.}~\bibnamefont {Agrawal}}, \bibinfo {author} {\bibfnamefont {A.~V.}\
  \bibnamefont {Dixit}}, \bibinfo {author} {\bibfnamefont {M.~H.}\ \bibnamefont
  {Awida}}, \bibinfo {author} {\bibfnamefont {A.~S.}\ \bibnamefont {Chou}},
  \bibinfo {author} {\bibfnamefont {M.}~\bibnamefont {Hollister}}, \bibinfo
  {author} {\bibfnamefont {S.}~\bibnamefont {Knirck}}, \bibinfo {author}
  {\bibfnamefont {A.}~\bibnamefont {Sonnenschein}}, \bibinfo {author}
  {\bibfnamefont {W.}~\bibnamefont {Wester}}, \bibinfo {author} {\bibfnamefont
  {J.~R.}\ \bibnamefont {Gleason}}, \bibinfo {author} {\bibfnamefont {A.~T.}\
  \bibnamefont {Hipp}}, \bibinfo {author} {\bibfnamefont {S.}~\bibnamefont
  {Jois}}, \bibinfo {author} {\bibfnamefont {P.}~\bibnamefont {Sikivie}},
  \bibinfo {author} {\bibfnamefont {N.~S.}\ \bibnamefont {Sullivan}}, \bibinfo
  {author} {\bibfnamefont {D.~B.}\ \bibnamefont {Tanner}}, \bibinfo {author}
  {\bibfnamefont {E.}~\bibnamefont {Lentz}}, \bibinfo {author} {\bibfnamefont
  {R.}~\bibnamefont {Khatiwada}}, \bibinfo {author} {\bibfnamefont
  {G.}~\bibnamefont {Carosi}}, \bibinfo {author} {\bibfnamefont
  {N.}~\bibnamefont {Robertson}}, \bibinfo {author} {\bibfnamefont
  {N.}~\bibnamefont {Woollett}}, \bibinfo {author} {\bibfnamefont {L.~D.}\
  \bibnamefont {Duffy}}, \bibinfo {author} {\bibfnamefont {C.}~\bibnamefont
  {Boutan}}, \bibinfo {author} {\bibfnamefont {M.}~\bibnamefont {Jones}},
  \bibinfo {author} {\bibfnamefont {B.~H.}\ \bibnamefont {LaRoque}}, \bibinfo
  {author} {\bibfnamefont {N.~S.}\ \bibnamefont {Oblath}}, \bibinfo {author}
  {\bibfnamefont {M.~S.}\ \bibnamefont {Taubman}}, \bibinfo {author}
  {\bibfnamefont {E.~J.}\ \bibnamefont {Daw}}, \bibinfo {author} {\bibfnamefont
  {M.~G.}\ \bibnamefont {Perry}}, \bibinfo {author} {\bibfnamefont {J.~H.}\
  \bibnamefont {Buckley}}, \bibinfo {author} {\bibfnamefont {C.}~\bibnamefont
  {Gaikwad}}, \bibinfo {author} {\bibfnamefont {J.}~\bibnamefont {Hoffman}},
  \bibinfo {author} {\bibfnamefont {K.~W.}\ \bibnamefont {Murch}}, \bibinfo
  {author} {\bibfnamefont {M.}~\bibnamefont {Goryachev}}, \bibinfo {author}
  {\bibfnamefont {B.~T.}\ \bibnamefont {McAllister}}, \bibinfo {author}
  {\bibfnamefont {A.}~\bibnamefont {Quiskamp}}, \bibinfo {author}
  {\bibfnamefont {C.}~\bibnamefont {Thomson}},\ and\ \bibinfo {author}
  {\bibfnamefont {M.~E.}\ \bibnamefont {Tobar}} (\bibinfo {collaboration} {ADMX
  Collaboration}),\ }\href {https://doi.org/10.1103/PhysRevLett.127.261803}
  {\bibfield  {journal} {\bibinfo  {journal} {Phys. Rev. Lett.}\ }\textbf
  {\bibinfo {volume} {127}},\ \bibinfo {pages} {261803} (\bibinfo {year}
  {2021})}\BibitemShut {NoStop}%
\bibitem [{\citenamefont {Fermilab}(2020)}]{SQMS}%
  \BibitemOpen
  \bibfield  {author} {\bibinfo {author} {\bibnamefont {Fermilab}},\ }\href
  {https://sqms.fnal.gov/research/} {\bibinfo {title} {Superconducting quantum
  materials and systems}} (\bibinfo {year} {2020})\BibitemShut {NoStop}%
\bibitem [{\citenamefont {Russenschuck}\ and\ \citenamefont
  {G.Vandoni}(2002)}]{CAS02}%
  \BibitemOpen
  \bibfield  {author} {\bibinfo {author} {\bibfnamefont {S.}~\bibnamefont
  {Russenschuck}}\ and\ \bibinfo {author} {\bibnamefont {G.Vandoni}},\
  }\href@noop {} {\bibinfo {title} {Cern accelerator school: Superconductivity
  and cryogenics for accelerators and detectors}} (\bibinfo {year}
  {2002})\BibitemShut {NoStop}%
\bibitem [{\citenamefont {Peccei}\ and\ \citenamefont
  {Quinn}(1977)}]{Peccei:1977hh}%
  \BibitemOpen
  \bibfield  {author} {\bibinfo {author} {\bibfnamefont {R.~D.}\ \bibnamefont
  {Peccei}}\ and\ \bibinfo {author} {\bibfnamefont {H.~R.}\ \bibnamefont
  {Quinn}},\ }\href {https://doi.org/10.1103/PhysRevLett.38.1440} {\bibfield
  {journal} {\bibinfo  {journal} {Phys. Rev. Lett.}\ }\textbf {\bibinfo
  {volume} {38}},\ \bibinfo {pages} {1440} (\bibinfo {year}
  {1977})}\BibitemShut {NoStop}%
\bibitem [{\citenamefont {Weinberg}(1978)}]{Weinberg:1977ma}%
  \BibitemOpen
  \bibfield  {author} {\bibinfo {author} {\bibfnamefont {S.}~\bibnamefont
  {Weinberg}},\ }\href {https://doi.org/10.1103/PhysRevLett.40.223} {\bibfield
  {journal} {\bibinfo  {journal} {Phys. Rev. Lett.}\ }\textbf {\bibinfo
  {volume} {40}},\ \bibinfo {pages} {223} (\bibinfo {year} {1978})}\BibitemShut
  {NoStop}%
\bibitem [{\citenamefont {Wilczek}(1978)}]{Wilczek:1977pj}%
  \BibitemOpen
  \bibfield  {author} {\bibinfo {author} {\bibfnamefont {F.}~\bibnamefont
  {Wilczek}},\ }\href {https://doi.org/10.1103/PhysRevLett.40.279} {\bibfield
  {journal} {\bibinfo  {journal} {Phys. Rev. Lett.}\ }\textbf {\bibinfo
  {volume} {40}},\ \bibinfo {pages} {279} (\bibinfo {year} {1978})}\BibitemShut
  {NoStop}%
\bibitem [{\citenamefont {Cervantes}\ \emph {et~al.}(2022)\citenamefont
  {Cervantes}, \citenamefont {Braggio}, \citenamefont {Gioccone}, \citenamefont
  {Frolov}, \citenamefont {Grasselino}, \citenamefont {Harnik}, \citenamefont
  {Melnychuk}, \citenamefont {Pilipenko}, \citenamefont {Posen},\ and\
  \citenamefont {Romanenko}}]{RaphaelSRF}%
  \BibitemOpen
  \bibfield  {author} {\bibinfo {author} {\bibfnamefont {R.}~\bibnamefont
  {Cervantes}}, \bibinfo {author} {\bibfnamefont {C.}~\bibnamefont {Braggio}},
  \bibinfo {author} {\bibfnamefont {B.}~\bibnamefont {Gioccone}}, \bibinfo
  {author} {\bibfnamefont {D.}~\bibnamefont {Frolov}}, \bibinfo {author}
  {\bibfnamefont {A.}~\bibnamefont {Grasselino}}, \bibinfo {author}
  {\bibfnamefont {R.}~\bibnamefont {Harnik}}, \bibinfo {author} {\bibfnamefont
  {O.}~\bibnamefont {Melnychuk}}, \bibinfo {author} {\bibfnamefont
  {R.}~\bibnamefont {Pilipenko}}, \bibinfo {author} {\bibfnamefont
  {S.}~\bibnamefont {Posen}},\ and\ \bibinfo {author} {\bibfnamefont
  {A.}~\bibnamefont {Romanenko}},\ }\href {https://arxiv.org/abs/2208.03183}
  {\bibfield  {journal} {\bibinfo  {journal} {arXiv}\ } (\bibinfo {year}
  {2022})},\ \bibinfo {note} {arXiv:2208.03183}\BibitemShut {NoStop}%
\bibitem [{\citenamefont {Parks}(1969)}]{Parks}%
  \BibitemOpen
  \bibfield  {author} {\bibinfo {author} {\bibfnamefont {R.}~\bibnamefont
  {Parks}},\ }\href@noop {} {\bibinfo {title} {Superconductivity in two
  volumes}} (\bibinfo {year} {1969})\BibitemShut {NoStop}%
\bibitem [{\citenamefont {Tinkham}(1996)}]{Tinkham}%
  \BibitemOpen
  \bibfield  {author} {\bibinfo {author} {\bibfnamefont {M.}~\bibnamefont
  {Tinkham}},\ }\href@noop {} {\bibinfo {title} {Introduction to
  superconductivity}} (\bibinfo {year} {1996})\BibitemShut {NoStop}%
\bibitem [{\citenamefont {Sikivie}(1985)}]{Sikivie1985}%
  \BibitemOpen
  \bibfield  {author} {\bibinfo {author} {\bibfnamefont {P.}~\bibnamefont
  {Sikivie}},\ }\href {https://doi.org/10.1103/PhysRevD.32.2988} {\bibfield
  {journal} {\bibinfo  {journal} {Phys. Rev. D}\ }\textbf {\bibinfo {volume}
  {32}},\ \bibinfo {pages} {2988} (\bibinfo {year} {1985})}\BibitemShut
  {NoStop}%
\bibitem [{\citenamefont {Rohlf}(1994)}]{Rohlf}%
  \BibitemOpen
  \bibfield  {author} {\bibinfo {author} {\bibfnamefont {J.~W.}\ \bibnamefont
  {Rohlf}},\ }\href@noop {} {\bibinfo {title} {Modern physics from a to z0}}
  (\bibinfo {year} {1994}),\ \bibinfo {note} {type I superconductors critical
  fields and temperatures}\BibitemShut {NoStop}%
\bibitem [{\citenamefont {Blatt}(1992)}]{Blatt}%
  \BibitemOpen
  \bibfield  {author} {\bibinfo {author} {\bibfnamefont {F.~J.}\ \bibnamefont
  {Blatt}},\ }\href@noop {} {\bibinfo {title} {Modern physics}} (\bibinfo
  {year} {1992}),\ \bibinfo {note} {type II superconductors critical fields and
  temperatures}\BibitemShut {NoStop}%
\bibitem [{\citenamefont {Jackson}(1975)}]{Jackson}%
  \BibitemOpen
  \bibfield  {author} {\bibinfo {author} {\bibfnamefont {J.~D.}\ \bibnamefont
  {Jackson}},\ }\href@noop {} {\emph {\bibinfo {title} {{Classical
  Electrodynamics; 2nd ed.}}}}\ (\bibinfo  {publisher} {Wiley},\ \bibinfo
  {address} {New York, NY},\ \bibinfo {year} {1975})\BibitemShut {NoStop}%
\bibitem [{\citenamefont {Sikivie}(1983)}]{Sikivie:1983ip}%
  \BibitemOpen
  \bibfield  {author} {\bibinfo {author} {\bibfnamefont {P.}~\bibnamefont
  {Sikivie}},\ }\href {https://doi.org/10.1103/PhysRevLett.51.1415} {\bibfield
  {journal} {\bibinfo  {journal} {Phys. Rev. Lett.}\ }\textbf {\bibinfo
  {volume} {51}},\ \bibinfo {pages} {1415} (\bibinfo {year}
  {1983})}\BibitemShut {NoStop}%
\bibitem [{\citenamefont {Asztalos}\ \emph {et~al.}(2010)\citenamefont
  {Asztalos}, \citenamefont {Carosi}, \citenamefont {Hagmann}, \citenamefont
  {Kinion}, \citenamefont {van Bibber}, \citenamefont {Hotz}, \citenamefont
  {Rosenberg}, \citenamefont {Rybka}, \citenamefont {Hoskins}, \citenamefont
  {Hwang}, \citenamefont {Sikivie}, \citenamefont {Tanner}, \citenamefont
  {Bradley},\ and\ \citenamefont {Clarke}}]{Asztalos:2009yp}%
  \BibitemOpen
  \bibfield  {author} {\bibinfo {author} {\bibfnamefont {S.~J.}\ \bibnamefont
  {Asztalos}}, \bibinfo {author} {\bibfnamefont {G.}~\bibnamefont {Carosi}},
  \bibinfo {author} {\bibfnamefont {C.}~\bibnamefont {Hagmann}}, \bibinfo
  {author} {\bibfnamefont {D.}~\bibnamefont {Kinion}}, \bibinfo {author}
  {\bibfnamefont {K.}~\bibnamefont {van Bibber}}, \bibinfo {author}
  {\bibfnamefont {M.}~\bibnamefont {Hotz}}, \bibinfo {author} {\bibfnamefont
  {L.~J.}\ \bibnamefont {Rosenberg}}, \bibinfo {author} {\bibfnamefont
  {G.}~\bibnamefont {Rybka}}, \bibinfo {author} {\bibfnamefont
  {J.}~\bibnamefont {Hoskins}}, \bibinfo {author} {\bibfnamefont
  {J.}~\bibnamefont {Hwang}}, \bibinfo {author} {\bibfnamefont
  {P.}~\bibnamefont {Sikivie}}, \bibinfo {author} {\bibfnamefont {D.~B.}\
  \bibnamefont {Tanner}}, \bibinfo {author} {\bibfnamefont {R.}~\bibnamefont
  {Bradley}},\ and\ \bibinfo {author} {\bibfnamefont {J.}~\bibnamefont
  {Clarke}},\ }\href {https://doi.org/10.1103/PhysRevLett.104.041301}
  {\bibfield  {journal} {\bibinfo  {journal} {Phys. Rev. Lett.}\ }\textbf
  {\bibinfo {volume} {104}},\ \bibinfo {pages} {041301} (\bibinfo {year}
  {2010})}\BibitemShut {NoStop}%
\bibitem [{\citenamefont {Asztalos}\ \emph {et~al.}(2011)\citenamefont
  {Asztalos}, \citenamefont {Carosi}, \citenamefont {Hagmann}, \citenamefont
  {Kinion}, \citenamefont {van Bibber}, \citenamefont {Hotz}, \citenamefont
  {Rosenberg}, \citenamefont {Rybka}, \citenamefont {Wagner}, \citenamefont
  {Hoskins}, \citenamefont {Martin}, \citenamefont {Sullivan}, \citenamefont
  {Tanner}, \citenamefont {Bradley},\ and\ \citenamefont
  {Clarke}}]{ASZTALOS201139}%
  \BibitemOpen
  \bibfield  {author} {\bibinfo {author} {\bibfnamefont {S.}~\bibnamefont
  {Asztalos}}, \bibinfo {author} {\bibfnamefont {G.}~\bibnamefont {Carosi}},
  \bibinfo {author} {\bibfnamefont {C.}~\bibnamefont {Hagmann}}, \bibinfo
  {author} {\bibfnamefont {D.}~\bibnamefont {Kinion}}, \bibinfo {author}
  {\bibfnamefont {K.}~\bibnamefont {van Bibber}}, \bibinfo {author}
  {\bibfnamefont {M.}~\bibnamefont {Hotz}}, \bibinfo {author} {\bibfnamefont
  {L.~J.}\ \bibnamefont {Rosenberg}}, \bibinfo {author} {\bibfnamefont
  {G.}~\bibnamefont {Rybka}}, \bibinfo {author} {\bibfnamefont
  {A.}~\bibnamefont {Wagner}}, \bibinfo {author} {\bibfnamefont
  {J.}~\bibnamefont {Hoskins}}, \bibinfo {author} {\bibfnamefont
  {C.}~\bibnamefont {Martin}}, \bibinfo {author} {\bibfnamefont {N.~S.}\
  \bibnamefont {Sullivan}}, \bibinfo {author} {\bibfnamefont {D.~B.}\
  \bibnamefont {Tanner}}, \bibinfo {author} {\bibfnamefont {R.}~\bibnamefont
  {Bradley}},\ and\ \bibinfo {author} {\bibfnamefont {J.}~\bibnamefont
  {Clarke}},\ }\href
  {https://doi.org/https://doi.org/10.1016/j.nima.2011.07.019} {\bibfield
  {journal} {\bibinfo  {journal} {Nucl. Instrum. Methods Phys. Res. A}\
  }\textbf {\bibinfo {volume} {656}},\ \bibinfo {pages} {39 } (\bibinfo {year}
  {2011})}\BibitemShut {NoStop}%
\bibitem [{\citenamefont {Stern}\ \emph {et~al.}(2015)\citenamefont {Stern},
  \citenamefont {Chisholm}, \citenamefont {Hoskins}, \citenamefont {Sikivie},
  \citenamefont {Sullivan}, \citenamefont {Tanner}, \citenamefont {Carosi},\
  and\ \citenamefont {van Bibber}}]{Stern}%
  \BibitemOpen
  \bibfield  {author} {\bibinfo {author} {\bibfnamefont {I.}~\bibnamefont
  {Stern}}, \bibinfo {author} {\bibfnamefont {A.~A.}\ \bibnamefont {Chisholm}},
  \bibinfo {author} {\bibfnamefont {J.}~\bibnamefont {Hoskins}}, \bibinfo
  {author} {\bibfnamefont {P.}~\bibnamefont {Sikivie}}, \bibinfo {author}
  {\bibfnamefont {N.~S.}\ \bibnamefont {Sullivan}}, \bibinfo {author}
  {\bibfnamefont {D.~B.}\ \bibnamefont {Tanner}}, \bibinfo {author}
  {\bibfnamefont {G.}~\bibnamefont {Carosi}},\ and\ \bibinfo {author}
  {\bibfnamefont {K.}~\bibnamefont {van Bibber}},\ }\href
  {https://doi.org/10.1063/1.4938164} {\bibfield  {journal} {\bibinfo
  {journal} {Review of Scientific Instruments}\ }\textbf {\bibinfo {volume}
  {86}},\ \bibinfo {pages} {123305} (\bibinfo {year} {2015})},\ \Eprint
  {https://arxiv.org/abs/https://doi.org/10.1063/1.4938164}
  {https://doi.org/10.1063/1.4938164} \BibitemShut {NoStop}%
\bibitem [{\citenamefont {Peng}\ \emph {et~al.}(2000)\citenamefont {Peng},
  \citenamefont {Asztalos}, \citenamefont {Daw}, \citenamefont {Golubev},
  \citenamefont {Hagmann}, \citenamefont {Kinion}, \citenamefont {LaVeigne},
  \citenamefont {Moltz}, \citenamefont {Nezrick}, \citenamefont {Powell} \emph
  {et~al.}}]{Peng2000569}%
  \BibitemOpen
  \bibfield  {author} {\bibinfo {author} {\bibfnamefont {H.}~\bibnamefont
  {Peng}}, \bibinfo {author} {\bibfnamefont {S.}~\bibnamefont {Asztalos}},
  \bibinfo {author} {\bibfnamefont {E.}~\bibnamefont {Daw}}, \bibinfo {author}
  {\bibfnamefont {N.}~\bibnamefont {Golubev}}, \bibinfo {author} {\bibfnamefont
  {C.}~\bibnamefont {Hagmann}}, \bibinfo {author} {\bibfnamefont
  {D.}~\bibnamefont {Kinion}}, \bibinfo {author} {\bibfnamefont
  {J.}~\bibnamefont {LaVeigne}}, \bibinfo {author} {\bibfnamefont
  {D.}~\bibnamefont {Moltz}}, \bibinfo {author} {\bibfnamefont
  {F.}~\bibnamefont {Nezrick}}, \bibinfo {author} {\bibfnamefont
  {J.}~\bibnamefont {Powell}}, \emph {et~al.},\ }\href
  {https://doi.org/http://dx.doi.org/10.1016/S0168-9002(99)00971-7} {\bibfield
  {journal} {\bibinfo  {journal} {Nuclear Instruments and Methods in Physics
  Research Section A: Accelerators, Spectrometers, Detectors and Associated
  Equipment}\ }\textbf {\bibinfo {volume} {444}},\ \bibinfo {pages} {569 }
  (\bibinfo {year} {2000})}\BibitemShut {NoStop}%
\bibitem [{\citenamefont {Gittleman}\ and\ \citenamefont
  {Rosenblum}(1966)}]{GittlemanRosenblum}%
  \BibitemOpen
  \bibfield  {author} {\bibinfo {author} {\bibfnamefont {J.~I.}\ \bibnamefont
  {Gittleman}}\ and\ \bibinfo {author} {\bibfnamefont {B.}~\bibnamefont
  {Rosenblum}},\ }\href@noop {} {\bibfield  {journal} {\bibinfo  {journal}
  {Physics Review Letters}\ }\textbf {\bibinfo {volume} {16}},\ \bibinfo
  {pages} {734} (\bibinfo {year} {1966})}\BibitemShut {NoStop}%
\bibitem [{\citenamefont {Pompeo}\ and\ \citenamefont {Silva}(2008)}]{Pompeo}%
  \BibitemOpen
  \bibfield  {author} {\bibinfo {author} {\bibfnamefont {N.}~\bibnamefont
  {Pompeo}}\ and\ \bibinfo {author} {\bibfnamefont {E.}~\bibnamefont {Silva}},\
  }\href@noop {} {\bibfield  {journal} {\bibinfo  {journal} {Physics Review B}\
  }\textbf {\bibinfo {volume} {78}},\ \bibinfo {pages} {094503} (\bibinfo
  {year} {2008})}\BibitemShut {NoStop}%
\bibitem [{\citenamefont {Alesini}\ \emph {et~al.}(2019)\citenamefont
  {Alesini}, \citenamefont {Braggio}, \citenamefont {Carugno}, \citenamefont
  {Crescini}, \citenamefont {D’Agostino}, \citenamefont {Gioacchino},\ and\
  \citenamefont {Tocci}}]{QUAX}%
  \BibitemOpen
  \bibfield  {author} {\bibinfo {author} {\bibfnamefont {D.}~\bibnamefont
  {Alesini}}, \bibinfo {author} {\bibfnamefont {C.}~\bibnamefont {Braggio}},
  \bibinfo {author} {\bibfnamefont {G.}~\bibnamefont {Carugno}}, \bibinfo
  {author} {\bibfnamefont {N.}~\bibnamefont {Crescini}}, \bibinfo {author}
  {\bibfnamefont {D.}~\bibnamefont {D’Agostino}}, \bibinfo {author}
  {\bibfnamefont {D.~D.}\ \bibnamefont {Gioacchino}},\ and\ \bibinfo {author}
  {\bibfnamefont {S.}~\bibnamefont {Tocci}},\ }\href@noop {} {\bibfield
  {journal} {\bibinfo  {journal} {Physics Review D}\ }\textbf {\bibinfo
  {volume} {99}},\ \bibinfo {pages} {101101} (\bibinfo {year}
  {2019})}\BibitemShut {NoStop}%
\bibitem [{\citenamefont {Ahn}\ \emph {et~al.}(2019)\citenamefont {Ahn},
  \citenamefont {Youm}, \citenamefont {Kwon}, \citenamefont {Chung},\ and\
  \citenamefont {K.~Semertzidis}}]{CAPPYBCO}%
  \BibitemOpen
  \bibfield  {author} {\bibinfo {author} {\bibfnamefont {D.}~\bibnamefont
  {Ahn}}, \bibinfo {author} {\bibfnamefont {D.}~\bibnamefont {Youm}}, \bibinfo
  {author} {\bibfnamefont {O.}~\bibnamefont {Kwon}}, \bibinfo {author}
  {\bibfnamefont {W.}~\bibnamefont {Chung}},\ and\ \bibinfo {author}
  {\bibfnamefont {Y.}~\bibnamefont {K.~Semertzidis}},\ }\href
  {https://arxiv.org/abs/1902.04551} {\bibfield  {journal} {\bibinfo  {journal}
  {arXiv}\ } (\bibinfo {year} {2019})},\ \bibinfo {note}
  {arXiv:1902.04551}\BibitemShut {NoStop}%
\bibitem [{\citenamefont {Posen}\ \emph {et~al.}(2022)\citenamefont {Posen},
  \citenamefont {Checchin}, \citenamefont {Melnychuk}, \citenamefont {Ring},\
  and\ \citenamefont {Gonin}}]{PosenNbSN}%
  \BibitemOpen
  \bibfield  {author} {\bibinfo {author} {\bibfnamefont {S.}~\bibnamefont
  {Posen}}, \bibinfo {author} {\bibfnamefont {M.}~\bibnamefont {Checchin}},
  \bibinfo {author} {\bibfnamefont {O.}~\bibnamefont {Melnychuk}}, \bibinfo
  {author} {\bibfnamefont {T.}~\bibnamefont {Ring}},\ and\ \bibinfo {author}
  {\bibfnamefont {I.}~\bibnamefont {Gonin}},\ }\href
  {https://arxiv.org/abs/2201.10733} {\bibfield  {journal} {\bibinfo  {journal}
  {arXiv}\ } (\bibinfo {year} {2022})},\ \bibinfo {note}
  {arXiv:2201.10733}\BibitemShut {NoStop}%
\bibitem [{\citenamefont {Brubaker}(2017)}]{Brubaker}%
  \BibitemOpen
  \bibfield  {author} {\bibinfo {author} {\bibfnamefont {B.~M.}\ \bibnamefont
  {Brubaker}},\ }\emph {\bibinfo {title} {First Results from the HAYSTAC axion
  search}},\ \href {https://arxiv.org/pdf/1801.00835.pdf} {Ph.D. thesis},\
  \bibinfo  {school} {Yale University} (\bibinfo {year} {2017})\BibitemShut
  {NoStop}%
\bibitem [{Sci(2022)}]{SciPy}%
  \BibitemOpen
  \href
  {https://docs.scipy.org/doc/scipy/reference/generated/scipy.optimize.curve_fit.html}
  {\bibinfo {title} {Reference manual for scipy.optimize.curve\_fit}} (\bibinfo
  {year} {2008-2022}),\ \bibinfo {note} {accessed: 2022-06-28}\BibitemShut
  {NoStop}%
\end{thebibliography}%

\end{document}